# IntLIM: Integration using Linear Models of metabolomics and gene expression data

Jalal K. Siddiqui[1], Elizabeth Baskin[1], Mingrui Liu[1], Carmen Z. Cantemir-Stone[1], Bofei Zhang[1], Russell Bonneville[2,3], Joseph P. McElroy[4], Kevin R. Coombes[1], Ewy A. Mathé[1,*]

[1]Department of Biomedical Informatics, College of Medicine, The Ohio State University, Columbus, Ohio, United States of America

[2]Biomedical Sciences Graduate Program, The Ohio State University, Columbus, Ohio, United States of America

[3]Comprehensive Cancer Center, Department of Internal Medicine, The Ohio State University, Columbus, Ohio, United States of America

[4]Center for Biostatistics, The Ohio State University, Columbus, Ohio, United States of America

*Corresponding author

Email Addresses:
JKS: jalal.siddiqui@osumc.edu
EB: baskin.18@osu.edu
ML: liu.4800@osu.edu
CZC: cantemir-stone.1@osu.edu
BZ: zhang.5675@osu.edu
RB: russell.bonneville@osumc.edu
JPM: joseph.mcelroy@osumc.edu
KRC: coombes.3@osu.edu
EAM: ewy.mathe@osumc.edu

## Abstract

**Background**: Integration of transcriptomic and metabolomic data improves functional interpretation of disease-related metabolomic phenotypes, and facilitates discovery of putative metabolite biomarkers and gene targets. For this reason, these data are increasingly collected in large (> 100 participants) cohorts, thereby driving a need for the development of user-friendly and open-source methods/tools for their integration. Of note, clinical/translational studies typically provide snapshot (e.g. one time point) gene and metabolite profiles and, oftentimes, most metabolites measured are not identified. Thus, in these types of studies, pathway/network approaches that take into account the complexity of transcript-metabolite relationships may neither be applicable nor readily uncover novel relationships. With this in mind, we propose a simple linear modeling approach to capture disease-(or other phenotype) specific gene-metabolite associations, with the assumption that co-regulation patterns reflect functionally related genes and metabolites.

**Results**: The proposed linear model, metabolite ~ gene + phenotype + gene:phenotype, specifically evaluates whether gene-metabolite relationships differ by phenotype, by testing whether the relationship in one phenotype is significantly different from the relationship in another phenotype (via a statistical interaction gene:phenotype p-value). Statistical interaction p-values for all possible gene-metabolite pairs are computed and significant pairs are then clustered by the directionality of associations (e.g. strong positive association in one phenotype, strong negative association in another phenotype). We implemented our approach as an R package, IntLIM, which includes a user-friendly R Shiny web interface, thereby making the integrative analyses accessible to non-computational experts. We applied IntLIM to two previously published datasets, collected in the NCI-60 cancer cell lines and in human breast tumor and non-tumor tissue, for which transcriptomic and metabolomic data are available. We demonstrate that IntLIM captures relevant tumor-specific gene-metabolite associations involved in known cancer-related pathways, including glutamine metabolism. Using IntLIM, we also uncover biologically relevant novel relationships that could be further tested experimentally.

**Conclusions**: IntLIM provides a user-friendly, reproducible framework to integrate transcriptomic and metabolomic data and help interpret metabolomic data and uncover novel gene-metabolite relationships. The IntLIM R package is publicly available in GitHub (https://github.com/mathelab/IntLIM) and includes a user-friendly web application, vignettes, sample data and data/code to reproduce results.

## Keywords

Metabolomics, Transcriptomics, Linear Modeling, Integration

# Background

Metabolomics data is increasingly collected in human biospecimens to identify putative biomarkers in diseases such as cancer [1-6]. Metabolites (small molecules < 1500 Daltons) are ideal candidates for biomarker discovery because they directly reflect disease phenotype and downstream effects of post-translational modifications [6]. However, interpretation of metabolomics data, including understanding how metabolite levels are modulated, is challenging. Reasons for this challenge include the presence of many (hundreds) of unidentified metabolites when untargeted approaches are applied [7, 8], and the fact that metabolomics profiles generated in human biospecimens are typically 'snapshots' or time-averaged representations of a disease state. Despite these difficulties, analyzing metabolomics data in light of other omics information, such as the transcriptome, can help to functionally interpret metabolomics phenotypes [9-15]. *Data integration,* or the use of multiple sources of information or data to provide a better model and understand a biological system [16], offers the opportunity to combine metabolomics data with other omics data-sets (e.g. transcriptome). Measurement and integration of the transcriptome and metabolome in the same cells, samples, or individuals, are thus increasingly applied to elucidate mechanisms that drive diseases, and to uncover putative biomarkers (metabolites) and targets (genes).

Current approaches that integrate transcriptomic and metabolomic data can be broadly categorized as numerical or pathway/network based. Numerical approaches include multivariate analyses (e.g. logistic regression, principal component analysis, partial least squares) and correlation-based approaches (e.g. canonical correlations) [17-19]. Differential correlation or coexpression methods have also been developed to capture changes in relationships between conditions [20]. Open-source tools, including MixOmics [21, 22] and DiffCorr [23], are available for integrating data but generally require in-depth statistical knowledge for their use and may not be as accessible to non-computational experts. Of note, such numerical approaches typically do not capture the complex and indirect relationships between transcripts and metabolites. For example, non-linear reaction kinetics mechanisms, metabolite-metabolite connections that regulate metabolite levels, and post-translational modifications all contribute to the complexity of gene-metabolite relationships [24, 25]. To better capture these complex relationships, pathway or network based approaches can be applied. Open-source tools such as Metaboanalyst [26], INMEX [27], XCMS Online [28], Metabox [29], and IMPALA [30] integrate transcriptomic and metabolomics data at a pathway level. One caveat of these approaches is that they rely on curated pathways or reaction-level information (knowledge of which enzymes produce a given metabolite) [18]. Pathway approaches are thus limited to metabolites that are identified and that can be mapped to pathways, which represents a fraction of what can be measured. In fact, of the 114,100 metabolites in the Human Metabolome Database [31-33], only 18,558 are detected and quantified, and of those, only 3,115 (17%) map to KEGG pathways. Further, network approaches that attempt to study the complex many to many associations between genes and metabolites may not scale well when tens of thousands of gene-metabolite pairs are evaluated.

Importantly, previous studies have shown that functionally related genes and metabolites show coherent co-regulation patterns [20, 34, 35]. We make this functionality assumption here and propose a linear modeling approach for integrating metabolomics and transcriptomics data to identify phenotype-specific gene-metabolite relationships. Of note, typical numerical integration approaches uncover patterns of molecular features that are globally correlated or aim to predict phenotype [20]. However, these methods do not directly and statistically test whether associations between metabolites and gene expression differ by phenotype. This distinction is important because global associations between genes and metabolites may not only reflect one phenotype of interest, but could reflect other features (e.g., environment, histology). As for methods that uncover differentially correlated pairs between conditions [35], they either do not capture pairs of features that are correlated in one group and not correlated in another group, or they bin relationships into different types (e.g. positive correlation in one group, negative correlation in another group), thereby making it difficult to compare more than 2 phenotypes [20, 34, 35]. Further, these approaches are not implemented into user-friendly frameworks. Our approach is thus advantageous because it directly evaluates the relationship between genes and metabolites in the context of phenotype, it can easily incorporate potential covariates, and is applicable to categorical (>= 2 groups) or continuous phenotypes. Further, our approach is implemented as a publicly available R package IntLIM (Integration through Linear Modeling), available at our GitHub repository [36], which includes an R Shiny web interface making it user-friendly to non-computational experts. In the wake of increasing amounts of metabolomics and transcriptomic data generated, availability of open-source, user-friendly, and streamlined approaches is key for reproducibility. Using IntLIM, we evaluated phenotype-specific relationships between gene and metabolite levels measured in the NCI-60 cancer cell lines [10], and in tumor and adjacent non-tumor tissue of breast cancer patients [9]. We demonstrate that IntLIM is useful for uncovering known and novel gene-metabolite relationships (which would require further experimental validation).

## Methods

**NCI-60 Cell Line Data Pre-processing**

The NCI-60 cancer cell line metabolomics (Metabolon platform) and gene expression data (Affymetrix U133 microarray) were downloaded from the Developmental Therapeutics Program (National Cancer Institute) website [10, 37]. Metabolomics and gene expression data, available in 57 cell lines, were pre-processed and normalized according to the Metabolon and Affymetrix MAS5 algorithms [38, 39], respectively. The metabolomics data contains 353 metabolites, of which 198 are unidentified. Each cell line is measured in triplicates (technical replicates), except for A498 and A549/ATCC, which had 4 and 2 technical replicates, respectively. The median of coefficients of variation (CVs) within technical replicate samples was calculated for each metabolite to assess consistency of abundance measurements. Metabolites with CVs < 0.3 were removed (280 metabolites remaining), abundances were log2 transformed, and the average technical replicate value was calculated for each metabolite. Next, the number of imputed values was estimated for each metabolite. The standard imputation method used by Metabolon is to impute missing values for a given metabolite by the minimum value of that metabolite across all samples. Thus, for each metabolite, the number of samples with a value equal to the minimum value (for that metabolite across all samples) minus "1" (one of those values is the true minimum value and should be subtracted) was used as an estimate of the number of missing values per metabolite. Metabolites with more than 80% imputed values were filtered out resulting in 220 metabolites, 111 of which are unidentified. Probes from the Chiron Affymetrix U133 microarrays were mapped to genes using the Bioconductor Ensembl database hgu133.plus.db [40]. In cases where more than one probe was matched to a given gene, the probe with the highest mean expression across all samples was retained for analysis, resulting in 17,987 genes with available expression. Lastly, we removed the 10% (arbitrary cutoff) of the lowest expressing genes, resulting in a total of 16,188 genes. For the linear modeling analyses, 220 metabolites and 16,188 genes were input.

For the NCI-60 cell line data, the phenotypes compared were leukemia cell lines vs. breast/prostate/ovarian (BPO) cell lines. Because this dataset was used to develop our approach, we purposefully chose cells from cancers that are known to be highly different in terms of their molecular profiles (e.g. blood cancer vs. solid tumor). The breast, prostate, and ovarian cancer cell lines were grouped together because they share susceptibility loci [41] and our aim was to increase sample size.

**Breast Cancer Data Pre-Processing**

Normalized gene expression (Affymetrix Gene Chip Human Gene 1.0 ST Arrays) and metabolomics (Metabolon) data in tumor and adjacent non-tumor tissue of breast cancer patients are publicly available through the Gene Expression Omnibus (GSE37751) and the supplementary data of the original publication, respectively [9, 42]. The data was normalized using the Metabolon algorithm (metabolites) and RMA algorithm [43] (genes), as previously described [9]. Both gene and metabolite levels are available for 61 tumor and 47 adjacent non-tumor breast tissue. The metabolomics data consists of 536 metabolites (203 of which are unidentified) in tumor and non-tumor tissue. Metabolites with more than 80% imputed values were removed, resulting in 379 metabolites, 119 of which are unidentified. Probes from the gene expression data not mapping to a gene symbol (Human Gene 1.0 ST Arrays) were removed. Similar to the NCI-60 data pre-processing, the probe with the highest mean expression was used for analysis when multiple probes mapped to a single gene. This resulted in 20,254 genes measured in tumor and non-tumor tissue. After removing the 10% lowest expression genes, we analyzed 18,228 genes. With this breast cancer data, our aim was to compare gene-metabolite associations between tumor and non-tumor tissue. A total of 379 metabolites and 18,228 genes were used for this analysis.

**IntLIM: Integration through Linear Modeling Approach**

The linear model applied to integrate transcriptomic and metabolomic data is:

$$m = \beta_1 + \beta_2 g + \beta_3 p + \beta_4 (g:p) + \varepsilon \qquad (1)$$

where **"m"** and **"g"** are normalized (see data pre-processing above) and log2-transformed metabolite abundances and gene levels respectively, **"p"** is phenotype (e.g. cancer type, tumor vs. normal), **"(g:p)"** is the statistical interaction [44] between gene expression and phenotype, and **"ε"** is the error term that is assumed to be independent and normally distributed (ε = N(0, σ )). A statistically significant two-tailed p-value of the **"(g:p)"** interaction term indicates that the slope relating gene expression and metabolite abundance is different from one phenotype compared to the other. Through this model, we can identify gene-metabolite associations that are specific to a particular phenotype (Figure 1). This model has been applied to all possible gene-metabolite pairs including those involving unidentified metabolites in the publicly available NCI-60 cancer cell line data [10] as well as previously published data from a breast cancer study [9]. Two-tailed p-values are subsequently corrected for multiple comparisons using the method by Benjamini and Hochberg to control the false discovery rate (FDR) [45]. Gene-metabolite pairs with an FDR-adjusted interaction p-value less than 0.10 or 0.05 in the NCI-60 cell line and breast cancer data, respectively, were used to determine statistical significance. (Due to the larger sample size in the breast cancer data set and the much larger amount of significant gene-metabolite pairs, our threshold for significance was more stringent).

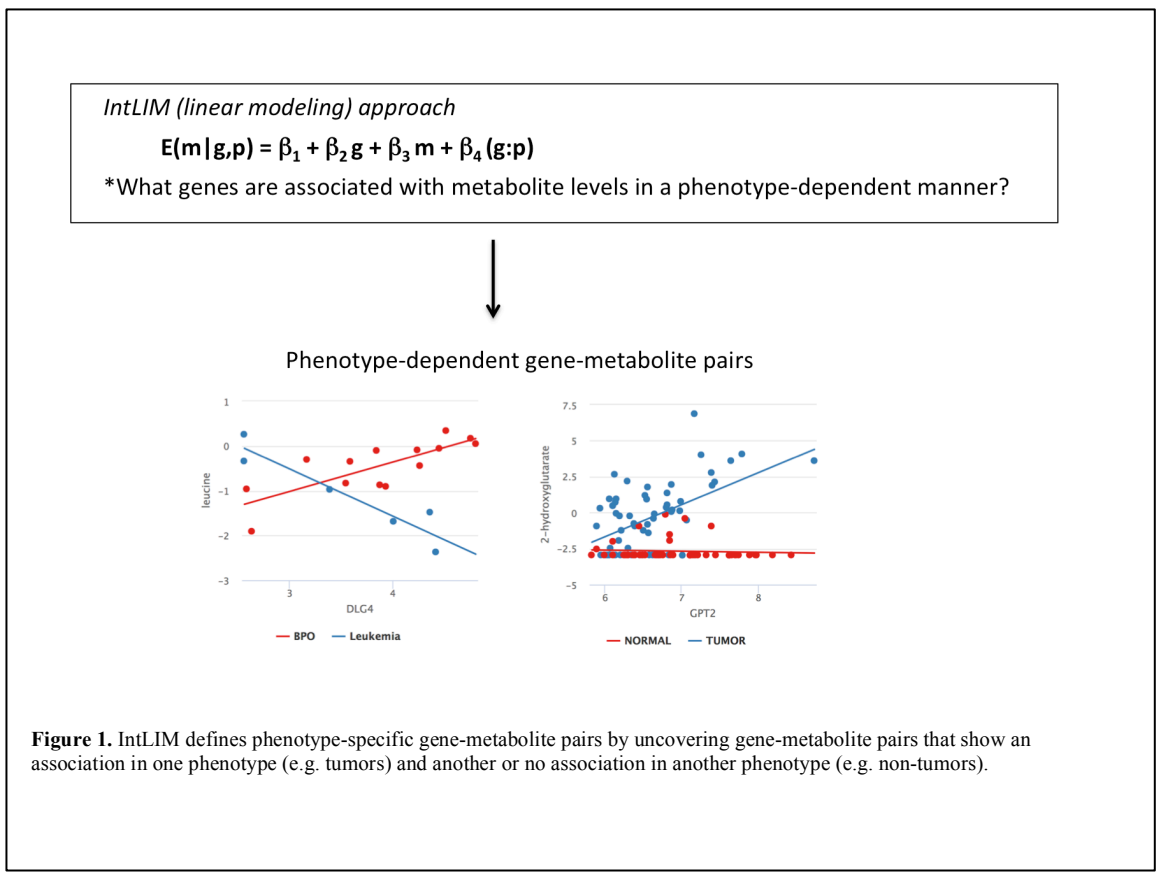

**Figure 1.** IntLIM defines phenotype-specific gene-metabolite pairs by uncovering gene-metabolite pairs that show an association in one phenotype (e.g. tumors) and another or no association in another phenotype (e.g. non-tumors).

To filter and cluster the list of statistically significant gene-metabolite pairs, the difference in Spearman correlations between the two phenotypic groups being compared (leukemia vs. BPO for NCI-60 cells and tumor vs. non-tumor for breast cancer tissue) was used as an effect size. Volcano plots of the difference in Spearman correlations vs. the –log10 (FDR-adjusted p-values) are depicted to visualize the distributions and help determine appropriate p-value and effect size cutoffs (Figure S3). For both datasets, a minimum absolute difference in correlations of 0.5 was used as an effect size cutoff.

The results can be visualized via a hierarchically clustered heatmap of gene-metabolite Spearman correlations calculated for each phenotypic group. Hierarchical clustering is performed with the *hclust* function. The Euclidean distance is used as the distance metric and the complete linkage method is used for agglomeration. The resulting dendrogram is used to create a heatmap that helps visualize how relevant gene-metabolite pairs cluster by their effect size (e.g. differences in Spearman correlation between the two phenotypic groups).

**IntLIM R Package**

A pipeline has been developed in the form of an R package to streamline integration of metabolomics and gene expression data using IntLIM. The package has been optimized and can solve a high number of linear models (3-7 million gene-metabolite pairs) in 2 to 6 minutes on a laptop with 2.7GHz quad-core Intel Core i7 processor and 16 GB, 2133MHz memory. Of note, IntLIM requires less than 3% of the time to solve all possible linear models compared to iterating through each model using the *lm* function in R for performing linear regression analysis as it contains a matrix algebra implementation of that function [46]. Extensive documentation is available in the package, including a vignette, and formatted NCI-60 and breast cancer datasets are linked and available in the IntLIM GitHub repository [36]. The steps for analysis are:

1) *Load data*: input CSV files containing normalized and log2-transformed gene expression data, normalized and log2-transformed metabolite abundance data, metadata for the samples (e.g. cancer status), and optionally metadata information on the genes and metabolites

2) *Filter data*: gene expression and metabolomics data are optionally filtered by gene and metabolite abundances and missing values

3) *Run IntLIM*: run linear models for all possible gene-metabolite pairs and extract FDR-adjusted interaction p-values and effect sizes (e.g. differences in slope/correlations between the groups)

4) *Filter gene-metabolite pairs*: filter results by user-input cutoffs of FDR-adjusted p-values and effect size. A volcano plot (absolute difference in correlation vs. –log10(FDR-adjusted p-values) is shown to help users determine appropriate adjusted p-value and effect size cutoffs. Resulting pairs are then clustered with hierarchical clustering, based on correlations within each, and visualized through heatmaps.

5) *Visualize relevant gene-metabolite pairs*: user-selected gene-metabolite pairs can be visualized through scatterplots, color-coded by phenotypic groups of interest (e.g. leukemia vs. BPO, tumor vs. non-tumor).

The IntLIM package also includes an RShiny web interface, a powerful tool that transforms complex analysis pipelines into interactive, user-friendly web applications [47]. The App guides users through all steps available in the package, as mentioned above. Of note, most plots are coded in highcharter [48] or plotly [49, 50] so users can promptly assess the effect of changing parameters on analysis results (e.g. immediate updates of tables and plots resulting from user changes of effect size and p-value cutoffs). We believe this interactivity accelerates data analysis and hence discovery of phenotype-specific gene-metabolite pairs. Further the app makes the analysis accessible to non-computational researchers. More information can be found in the S1 Appendix: IntLIM documentation.

**Pathway Analysis**

Pathway and upstream regulator analyses were performed using the Ingenuity Pathway Analysis (IPA) software. The list of genes or identified metabolites from each cluster (e.g. highly correlated in one group but no correlation in the other) of statistically significant gene-metabolite pairs were input to conduct pathway analysis to analyze input genes or metabolites in the context of biological pathways or functions [51]. IPA also includes an upstream regulator analysis to determine whether those molecules were associated with a particular upstream regulator. P-values, calculated from the Right-tailed Fisher's Exact Test, reflect whether the number of overlapping molecules associated with a particular pathway or upstream regulator is greater than expected by chance [52]. For upstream regulator analysis, both direct and indirect relationships between molecules and their targets were considered (confidence = Experimentally observed). [53].

**Results**

**IntLIM (Integration through LInear Modeling)**

Our goal is to find gene-metabolite pairs that have a strong association in one phenotype (e.g. leukemia vs. breast/prostate/ovarian cancers (BPO), tumor vs. non-tumor) and an inverse or no association in another phenotype. We anticipate that gene-metabolite relationships that are phenotype dependent will help interpret metabolomics phenotypes and will highlight molecular functions and pathways worth evaluating further. With accumulating transcriptomic and metabolomics data generated in the same samples, uncovering phenotype-specific relationships could elucidate novel co-regulation patterns. Because commonly leveraged untargeted metabolomics approaches produce large amounts of unidentified metabolites, approaches that rely on reaction-level or pathway annotations may not be sufficient to capture all or novel relationships. To accomplish our goal, we thus rely on numerical data integration and developed a linear modeling approach that predicts metabolite levels from gene expression in a phenotype-dependent manner (Figure 1) (see Methods). Unlike correlation-based and logistic regression approaches, our approach specifically evaluates whether the association between gene and metabolite levels is related to a phenotype. Furthermore, it is important to keep in mind that metabolite abundances can be modulated by a group of enzymes, which in turn are regulated by a myriad of regulatory processes (e.g. transcription, post-translational modifications). Thus, gene expression, protein abundances, and metabolite levels do not always have linear relationships. While these more complex relationships will not be readily detected using our approach [14], co-regulated gene-metabolite relationships tend to share biological functions [34] and we make this assumption here. Our approach is implemented as an R package, which is publicly available through GitHub (See Methods and IntLIM Documentation in S1 Appendix) [36].

**Application to NCI-60 Data**

The NCI-60 cell lines [10] were developed as a drug-screening tool focusing on a range of cancer types, including renal, colon, prostate, breast, ovarian, leukemia, and non-small cell lung cancer [54]. Transcriptomic (Affymetrix) and metabolomic (Metabolon) data are available for 57 of those cell lines [10] We applied IntLIM to identify cancer-type specific gene-metabolite associations. The two major subgroups compared were leukemia (6 cell lines: CCRF-CEM, HL-60 (TB), K-562, MOLT-4, RPMI-8226, SR) vs. the breast/prostate/ovarian (BPO) cancer cell lines (14 total cell lines: BT-549, DU-145, HS 578T, IGROV1, MCF7, MDA-MB-231/ATCC, NCI/ADR-RES, OVCAR-3, OVCAR-4, OVCAR-5, OVCAR-8, PC-3, SK-OV-3, T-47D) consisting of 16,188 genes and 220 metabolites (see Methods). The latter cancers were grouped together as they share common susceptibility loci [41]. Unsupervised clustering using principal components analysis (PCA) on the log2-transformed and filtered metabolomics and gene expression data (Figures S1A and B) clearly delineates the two major subgroups (Figure S1C and D).

All possible combinations of gene-metabolite pairs (3,561,360 models run) were evaluated, using "BPO" and "leukemia" as cancer type. We identified 1,009 cancer-type dependent gene-metabolite associations (FDR-adjusted p-value < 0.1 and correlation difference effect size > 0.5, Data S1, Figure S3A) involving 785 genes and 68 metabolites, of which 37 are unidentified. Clustering of these pairs by the direction of association (e.g. positive or negative correlation) within each cancer type subgroup revealed two major clusters (Figure 3). First, the "leukemia correlated cluster" consists of 545 gene-metabolite pairs (429 unique genes and 54 unique metabolites of which 31 are unidentified) with relatively high positive correlations in leukemia cell lines and low or negative correlations in BPO cell lines (Figure 2A). Second, the "leukemia anti-correlated cluster" consists of 464 gene-metabolite pairs (356 unique genes and 45 unique metabolites of which 24 are unidentified) with relatively high negative correlations in leukemia cell lines and positive or low negative correlations in BPO cell lines. Two of the top ranked gene-metabolite pairs (ranked in descending order of absolute value of Spearman correlation differences between BPO and leukemia) in the leukemia correlated and leukemia anti-correlated clusters are FSCN1-malic acid (Figure 2B) and DLG4-leucine (Figure 2C), respectively. FSCN1 and malic acid (Figure 2B) are positively correlated in leukemia (r=0.94) but negatively correlated in BPO cancers (r = -0.75) (Figure 2B). FSCN1 is associated with the progression of prostate cancer [55], while malic acid (or ionized malate) is an intermediate involved in glutamine metabolism pathways that play major roles in cancer metastasis [56, 57]. DLG4 and leucine (Figure 2C) are negatively correlated in leukemia (r = -0.92) but positively correlated (r = 0.78) in BPO cancers (Figure 2C). DLG4 is downregulated in human cervical cancer cell lines infected with human papillomavirus and may act as a tumor suppressor [58], while leucine deprivation inhibits cell proliferation and induces apoptosis in breast cancer cells [59]. Interestingly, leucine supplementation has been shown to enhance pancreatic cancer growth in mouse models [60]. These opposing correlations of DLG4-leucine and FSCN1-malic acid between leukemia and BPO suggest possible tissue-specific relationships that can be differentially targeted.

Pathway analysis on 419 unique and mappable genes in the "leukemia correlated cluster" showed enrichment of the following pathways: acute phase response signaling, 1D-myo-inositol hexakisphosphate biosynthesis, hepatic fibrosis/hepatic stellate cell activation, CDK5 signaling, and PAK signaling (Table S1). The "leukemia anti-correlated cluster" genes (N=351) were enriched for endothelial NOS signaling, CREB signaling in neurons, dTMP de novo biosynthesis, Huntington's Disease signaling, and the P2Y purigenic receptor signaling pathway (Table S1). Most of these pathways are relevant to cancer biology. For example, nitric oxide has been found to have both tumor suppressive (e.g. promoting apoptosis, inhibition of cancer

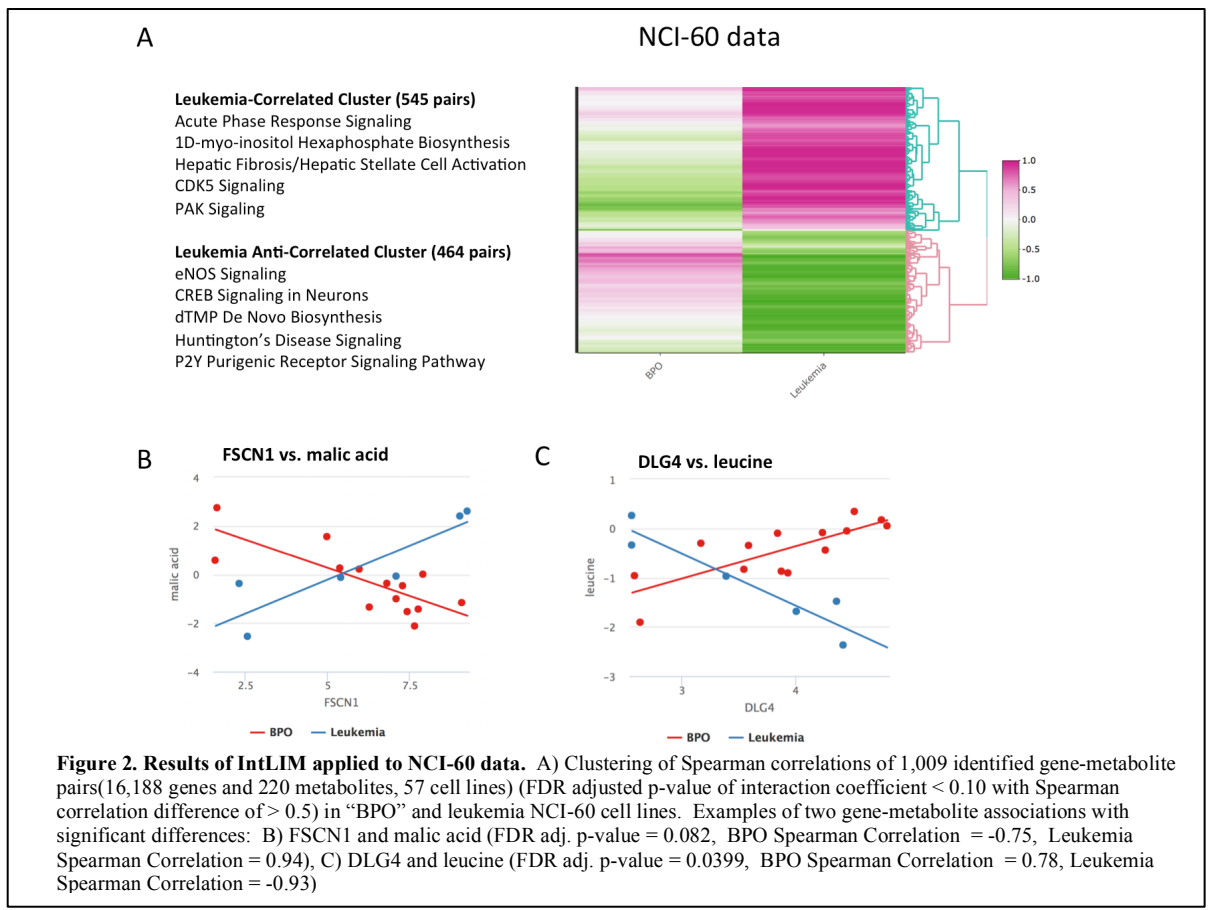

**Figure 2. Results of IntLIM applied to NCI-60 data.** A) Clustering of Spearman correlations of 1,009 identified gene-metabolite pairs(16,188 genes and 220 metabolites, 57 cell lines) (FDR adjusted p-value of interaction coefficient < 0.10 with Spearman correlation difference of > 0.5) in "BPO" and leukemia NCI-60 cell lines. Examples of two gene-metabolite associations with significant differences: B) FSCN1 and malic acid (FDR adj. p-value = 0.082, BPO Spearman Correlation = -0.75, Leukemia Spearman Correlation = 0.94), C) DLG4 and leucine (FDR adj. p-value = 0.0399, BPO Spearman Correlation = 0.78, Leukemia Spearman Correlation = -0.93)

growth) and tumor promoting properties (promotion of angiogenesis, DNA repair mechanisms) [61]. cAMP-regulator element binding protein (CREB) has been shown to be over-expressed and phosphorylated in several cancers (including acute myeloid leukemia) and might play a role in cancer pathogenesis [62]. These preliminary results demonstrate how different pathways may be differentially regulated in a cancer-type dependent manner. Since only 9 of 54 and 10 of 45 metabolites in the leukemia correlated and leukemia anti-correlated clusters, respectively, could be mapped to Human Metabolome Database (HMDB) IDs [31-33], pathway analyses were not possible for the metabolites.

**Application to Breast Cancer Data**

We further applied IntLIM to a previously published breast cancer study [9]. Gene expression and metabolomics profiling of tumor (n = 61) and adjacent non-tumor tissue samples (n = 47) was measured in tissue from breast cancer patients [9]. Importantly, gene expression and metabolomics were measured in the same tissue biospecimens. The original study identified a relationship between MYC activation and 2-hydroxyglutarate (2-HG) accumulation as associated with poor prognosis in breast cancer [9]. Studies involving MYC overexpression and knockdown in human mammary epithelial and breast cancer cells further corroborated this relationship [9]. When assessing the relationship between MYC gene expression and 2-HG though, we did not observe this association at the transcription level (Fig. 3C). Our goal was thus to identify other potential regulators of 2-hydroxyglutarate accumulation in breast cancer tissue, and to assess whether other gene-metabolite associations were specific to either tumor or non-tumor tissue. The data consists of 18,228 genes and 379 metabolites (119 unidentified) measured in 61 tumor samples and 47 adjacent non-tumor samples (Figure S2A and S2B). Unsupervised clustering of gene and metabolite abundances separated tumor from non-tumor tissue (Figure S2C and S2D).

IntLIM was applied to all possible combinations of gene-metabolite pairs (6,908,412 models), with tumor and non-tumor as the phenotype. Our approach identified 2,842 tumor-dependent gene-metabolite correlations (FDR-adjusted interaction p-value < 0.05, and a Spearman correlation difference > 0.5) involving 761 genes and 212 metabolites of which 48 are unidentified. (Data S2, Figure S3B). The resulting heatmap of gene-metabolite Spearman correlations for tumor and non-tumor groups is divided into two major clusters (Figure 3A). The first is a "tumor-correlated cluster" of 1,038 gene-metabolite pairs (288 unique genes and 155 metabolites of which 35 are unidentified) with relatively high correlations in tumor samples and mostly negative correlations in non-tumor samples. The second major cluster, "tumor anti-correlated cluster", comprises 1,804 gene-metabolite pairs (479 unique genes and 188 metabolites of which 39 are unidentified) with high negative correlations for tumor samples and mostly negative correlations for non-tumor samples.

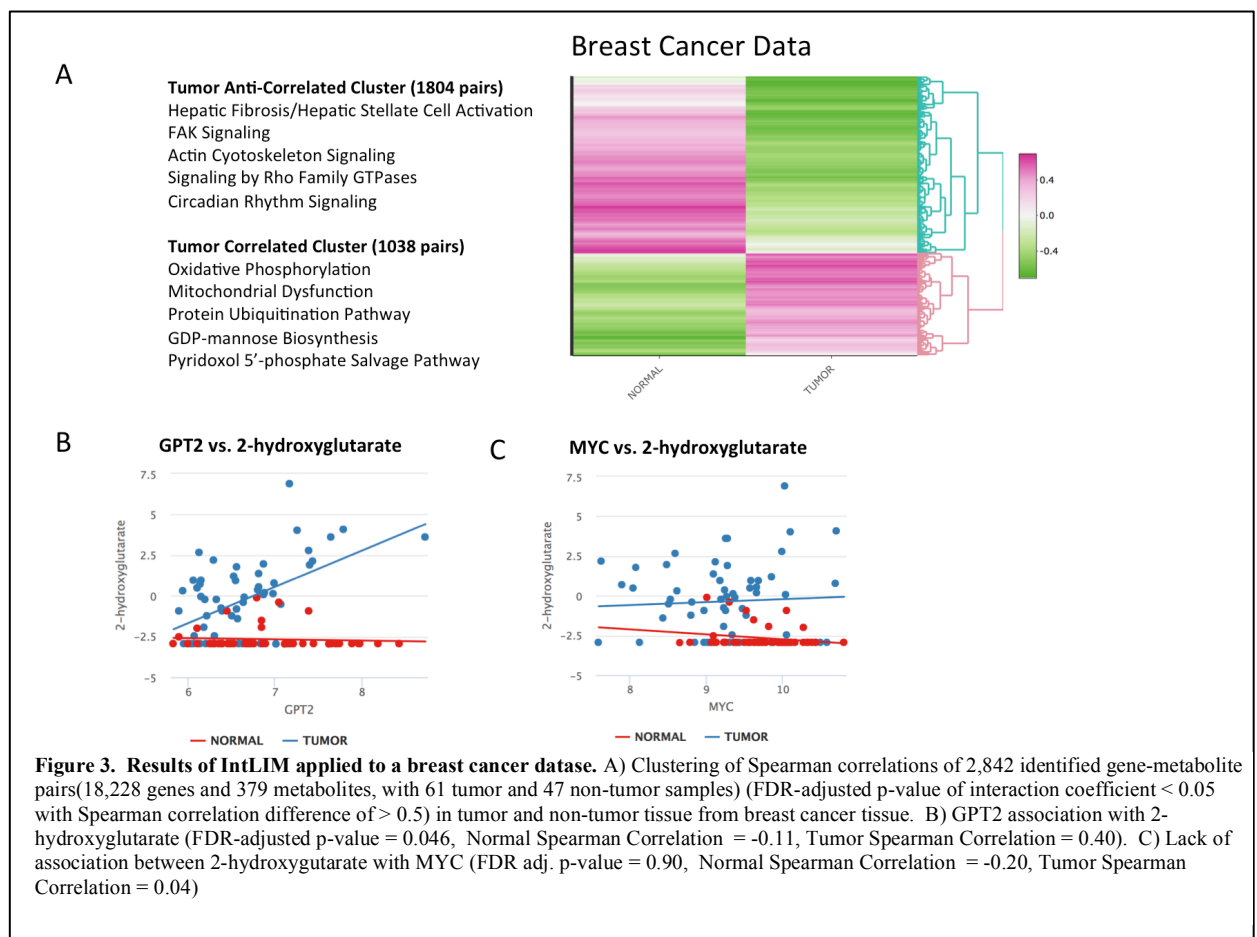

**Figure 3. Results of IntLIM applied to a breast cancer datase.** A) Clustering of Spearman correlations of 2,842 identified gene-metabolite pairs(18,228 genes and 379 metabolites, with 61 tumor and 47 non-tumor samples) (FDR-adjusted p-value of interaction coefficient < 0.05 with Spearman correlation difference of > 0.5) in tumor and non-tumor tissue from breast cancer tissue. B) GPT2 association with 2-hydroxyglutarate (FDR-adjusted p-value = 0.046, Normal Spearman Correlation = -0.11, Tumor Spearman Correlation = 0.40). C) Lack of association between 2-hydroxygutarate with MYC (FDR adj. p-value = 0.90, Normal Spearman Correlation = -0.20, Tumor Spearman Correlation = 0.04)

Upstream analysis of the genes involved in the tumor-correlated cluster (N = 283) did identify MYC as an upstream transcriptional regulator (Right-tailed Fisher's Exact Test p-value = $6 \times 10^{-3}$), even though MYC and 2-HG are not differentially associated (Figure 3C). 2-HG was, however, found to be associated with GPT2 (FDR adj p-value = 0.046, r = 0.40 in tumors, and r=-0.11 in non-tumors) (Figure 3B, Data S2). GPT2 plays a role in glutamine metabolism and encodes a glutamic-pyruvic transaminase that catalyzes reverse transamination between alanine and 2-oxoglutarate to generate pyruvate and glutamate [63]. Cancer cells exhibit a metabolic reprogramming that results in increased lactate acid production in the Warburg effect and the use of glutamine to replenish the tricarboxylic acid cycle (TCA) [64, 65]. The role of GPT2 serves to drive the utilization of glutamine as a carbon source for TCA analplerosis [63, 65]. While the exact mechanisms underlying increased levels of 2-hydroxyglutarate in breast cancer cells are not all known, our results suggest that metabolic reprogramming changes the relationship between GPT2 and 2-hydroxyglutarate. Furthermore, GPT2 is found to be in 18 (FDR adjusted p-value < 0.05 and correlation difference > 0.5) other tumor-specific gene-metabolite associations (Data S2).

In addition to GPT2 and 2-HG, we identified 15 other gene-metabolite pairs involving metabolites linked to glutamine metabolism. Of those genes paired with glutamine, ASNS, which encodes asparagine synthetase, is directly involved in metabolizing glutamine [66] and

SLC7A1 is involved in glutamine transport [64] (Data S2). Furthermore, there are 65 gene-metabolite pairs with glutamate and 25 pairs involving alanine (Data S2), and 5 gene-metabolite pairs involving the WIF gene, which is part of the Wnt signaling pathway [9] (Data S2).

Pathway analysis revealed that genes in the "tumor-correlated cluster" (283 mapped into IPA out of 288 genes) were enriched for oxidative phosphorylation, mitochondrial dysfunction, protein ubiquitination pathway, GDP-mannose biosynthesis, and the pyridoxal 5'-phosphate salvage pathway (Table S2). Genes in the "tumor anti-correlated cluster" (468 mapped onto IPA out of 479 genes) were enriched for hepatic fibrosis/hepatic stellate cell activation, FAK signaling, actin cytoskeleton signaling, signaling by Rho family GTPases, and circadian rhythm signaling (Table S2). Expectedly, we find that pathways such as FAK signaling, actin cytoskeleton, the protein ubiquitination pathway, and circadian rhythm signaling have strong links to breast cancer pathogenesis [67-71]. Of note, the top two pathways in the tumor-correlated cluster (oxidative phosphorylation and mitochondrial dysfunction) play roles in cellular energetics [72].

Pathway analysis of the metabolites in the "tumor-correlated cluster" (100 mapped onto IPA out of 155 metabolites) resulted in enrichment of pathways related to tRNA charging and nucleotide degradation (Table S3). The "tumor anti-correlated cluster" (115 mapped onto IPA out of 188 metabolites) was also enriched for tRNA charging, citrulline metabolism, urea cycle, purine nucleotide degradation, and purine ribonucleosides degradation to ribose-1-phosphate (Table S3). Pathways related to tRNA and the urea cycle have been implicated in cancer [73-75]. Citrulline metabolism and the urea cycle have also been linked to glutamine metabolism [57, 76, 77]. These findings are consistent with previous studies [9, 57, 63, 64] that highlight the role of glutamine metabolism in cancer cell proliferation and maintenance, especially with regards to breast cancer[9]. Further, the urea cycle has been shown to be implicated in breast cancer and is linked to glutamine metabolism[77]. Notably, our IntLIM results identify 2 gene-metabolite pairs with urea and 5 gene-metabolite pairs with arginine (FDR-adjusted p-value of 0.05 or less, absolute Spearman Correlation difference > 0.5), a major metabolite in the urea cycle (S2 Data) [77].

## Discussion

As more and more transcriptomic and metabolomic data are collected in the same samples or individuals, there is a need for streamlined methods and associated user-friendly tools that integrate these data. We implemented a novel linear modeling approach into an IntLIM R package that includes a user-friendly web interface, to statistically test whether gene and metabolite associations differ by phenotype. Formally testing this dependency on phenotype differentiates our approach from other numerical integration approaches such as logistic regression and canonical correlations. Compared to other existing methods that take into account phenotype dependency [20, 34], IntLIM is user-friendly, it uses a well-developed methodology (linear model interactions), can easily account for other covariables (e.g. gender, BMI, etc.), and can be applied to phenotypes that have more than two categories or are continuous. Ultimately, uncovering phenotype-specific relationships can provide insight into how metabolites are being regulated by genes and on which pathways may be involved in these phenotype-specific changes.

While knowledge of relevant pathways is powerful in developing potential disease interventions and treatments, pathway enrichment analyses are hampered by the large fraction of metabolites that are identified or cannot be mapped to pathways.  Importantly, IntLIM uncovers phenotype-dependent gene-metabolite associations without *a priori* curated information on pathways and networks, allowing discovery of potentially novel associations (that would require further experimental validation). Because untargeted metabolomics data produces many unidentified features, phenotype-specific associations with IntLIM could help further characterize these unidentified molecules. These data-driven discoveries would require further experimental validation and could generate new hypothesis to be tested. When pathway annotations are available though, pathway enrichment analysis of genes and metabolites that show similar patterns (e.g. positive correlation in tumors but no correlation in non-tumors) can offer greater insight onto pathways that are altered between phenotypes. With this in mind, IntLIM produces a list of relevant genes and metabolites that could be input into pathway integration approaches and software [26, 28-30].

To demonstrate the utility of IntLIM to uncover cancer-relevant gene-metabolite relationships, we evaluated transcriptomic and metabolomics data measured in the NCI-60 cell lines [10] and breast tumor/adjacent non-tumor tissue [9] (Figures 2 and 3).  In both these data sets, we uncovered biologically relevant gene-metabolite relationships and pathways. For example, glutamine metabolism clearly stood out as an altered pathway in the breast cancer data, in line with previous published results [9]. Interestingly, we also uncovered novel putative associations, such as the possible modulation by GPT2 of 2-hydroxyglutarate accumulation in breast cancer tissue (validation of this relationships would require further experimentation).

While this first iteration of IntLIM uncovers phenotype-specific gene-metabolite pairs, the approach can easily be extended to other omics data (e.g., metabolomics/microbiome data, metabolomics/proteome, proteome/transcriptome).  Of note, because IntLIM makes use of a linear model, we assume that the independent variables (e.g. metabolite levels) are normally distributed to meet the normality assumption. We have verified the normality assumption in the NCI-60 and breast cancer datasets and leave it up to the user to appropriately transform and check the normality of their data prior to using IntLIM.  Furthermore, our current linear model does not make use of the fact that some of the samples may be paired.  In our breast cancer data [9], only a subset of the patients (N = 41) have both tumor and adjacent non-tumor available.  It would be feasible to take into consideration the paired nature of the samples using a mixed model methodology, and thereby increase our power to detect significant relationships. Finally, future developments of IntLIM will accommodate greater flexibility in defining models. For example, we will include the capability of testing whether phenotype-specific gene-metabolite associations are independent of other putative confounders (e.g. age, gender, race, etc).  Further, while IntLIM currently only supports a binary phenotype, it is readily generalizable to multicategorical phenotypes.

Like most approaches, IntLIM and the studies conducted are not without limitations. The biochemical pathways that drive gene expression to protein production to post-translational modifications to metabolite production/consumption are complex [24]. The abundance of a given metabolite typically depend on a group of enzymes that produce/consume that metabolite. Additionally, those enzymes have distinct kinetic parameters, and their activity depends on a range of posttranslational modifications and regulatory processes. As a result, transcript levels are not the only factors that modulate metabolite abundance, and the gene-metabolite relationship may not be linear. In this regard, IntLIM may not adequately capture these complex relationships. Nonetheless, linear-based approaches are well-developed, have successfully been applied when integrating omics data, and co-regulated genes and metabolites tend to be associated with functional roles [10, 20, 34]. Further, we demonstrate that this simple approach can identify biologically meaningful, putative phenotype-dependent gene-metabolite relationships that can be investigated with further experiments. Another limitation is that IntLIM does not take into consideration time-dependency of biochemical reaction steps, especially given the time delay between gene expression and protein production and further on metabolite production/consumption. However, in clinical and translational applications, metabolomic and transcriptomic data is typically collected at a "snapshot" in time, where time-dependent analyses are not possible [78]. Lastly, our approach, along with other numerical and pathway based integration approaches, does not take into account cellular heterogeneity in specimens analyzed, even though this heterogeneity could impact gene-metabolite correlations in different regions of cells or tissues [79]. Because IntLIM remains agnostic to the input, especially with regards to cell/tissue heterogeneity, it is the user's responsibility to interpret the data as well as design future experiments to test findings from results. Despite these limitations, IntLIM provides a user-friendly, reproducible framework to integrate metabolomics and transcriptomics data, or other omics data and provides a readily implementable first step in integration.

## Conclusions

Metabolomics and transcriptomic data are increasingly collected in the same samples to uncover putative metabolite biomarkers and gene therapeutic targets. User-friendly approaches that integrate these data types will thus facilitate data interpretation in these studies, and could generate data-driven hypothesis. With this in mind, we developed a novel linear modeling approach that statistically tests whether gene-metabolite associations are specific to particular phenotypes (tumor vs. non-tumor, cancer-type, etc.). Our approach is available as a publicly available R package, IntLIM, with an associated user-friendly web application. We applied IntLIM to two cancer datasets and uncovered known and novel gene-metabolite pairs and pathways that were associated with cancer phenotypes. It is our hope that IntLIM will assist researchers, with or without computational expertise, in formulating novel hypothesis and proposing new studies especially with regards to the gene-metabolite pairs identified. Integrating the results with pathway analysis tools will provide further insight. The

IntLIM R package and App are available for download via GitHub and a sample data-set and vignette are provided for users.

## Abbreviations

**BPO:** Breast/Prostate/Ovarian, **IntLIM:** Integration through Linear Modeling

## Declarations

### Ethics Approval and Consent to Participate

Not applicable

### Consent for Publication

Not applicable

### Availability of Data and Materials

The R package, including a vignette and sample data set, is available online on the Github repository: [https://github.com/mathelab/IntLIM/]. Formatted NCI-60 datasets are available at: [https://github.com/Mathelab/NCI60_GeneMetabolite_Data]. Formatted breast cancer datasets are available at: [https://github.com/Mathelab/BreastCancerAmbs_GeneMetabolite_Data].

### Competing Interests

The authors declare that they have no competing interests.

### Author's Contributions

JKS helped design study, conducted analyses, and developed the software. EB assisted with conducting analyses and design of software. ML helped with design and development of the software. CZC assisted with analyses of results. BZ assisted with developing the software. RB assisted with developing software and with conducting analyses. JPM and KRC helped analyze and interpret results and offered suggestions for manuscript. EM designed study, conducted analyses, and developed the software. All authors read and approved the final manuscript.

### Acknowledgements

We thank Drs. Chris Beecher and Stefan Ambs for helpful discussions regarding the quality control and analysis of the NCI-60 data and the breast cancer data, respectively.

**Funding**

This work was supported by funding from The Ohio State University Translational Data Analytics Institute and startup funds from The Ohio State University to Dr. Ewy Mathé. This work was also supported by the The Ohio State University Discovery Themes Foods for Health postdoctoral fellowship to Dr. Jalal Siddiqui.

# S1 Appendix

# IntLIM Documentation

## Summary

Interpretation of metabolomics data is very challenging. Yet it can be eased through integration of metabolmoics with other 'omics' data. The IntLIM package, which includes a user-friendly RShiny web interface, aims to integrate metabolomic data with transcriptomic data. We implement a simple linear modeling approach to integration, where we focus on understanding how specific gene-metabolite associations are affected by phenotypic features. To this end, we include an interaction term in our linear model that specifically tests whether a gene-metabolite association differs by phenotype. The overall workflow involves the following steps:

1) input gene expression/metabolomics data files,
2) filter data sets by gene and metabolite abundances and imputed values,
3) run linear models on all possible gene-metabolites pairs and extract FDR-adjusted interaction p-values,
4) filter results by FDR-adjusted p-values and Spearman correlation differences,
5) plot/visualize user-defined gene-metabolite associations

The package, source code, vignettes, and formatted datasets used in to produce results in this manuscript are available for download on GitHub (https://github.com/mathelab/IntLIM).

## Installing and Running IntLIM

### Installing IntLIM

The "devtools" package (1) is the simplest way to directly install IntLIM:

```
install.packages(devtools)
library(devtools)
install_github("mathelab/IntLIM")
```

In some cases, it may be necessary to install the Bioconductor package *MultiDataSet* (Hernandez-Ferrer et al, 2017):

```
## try http:// if https:// URLs are not supported
source("https://bioconductor.org/biocLite.R")
biocLite("MultiDataSet")
```

### Running IntLIM

The IntLIM package provides the necessary functions to carry out the workflow of integrating gene expression and metabolomics data and finding phenotype-dependent gene-metabolite associations. The information below is meant to provide an overview of the workflow and the functions available in the package. Details on usage and parameters available for each function are included in the package documentation provided by each function. This documentation is accessed by typing "?functionname", where "functionname" is the name of the function, such as ReadData, FilterData, etc.

**Step1: Read in Data**

Through the ReadData() function, users input a comma-separated-values (CSV) file that contains information on the input files. This input CSV file must contain 2 columns and 6 rows and must include the following:

> type,filenames
> metabData,myfilename
> geneData,myfilename
> metabMetaData,myfilename (optional)
> geneMetaData,myfilename (optional)
> sampleMetaData,myfilename

The "myfilename" represent file names for the respective data types without the file directory, which is assumed to be the same as the input file. **Thus, IntLIM assumes that all input files are in the same directory.** The file names for the gene data ('geneData') and metabolite data ('metabData') must be normalized and can be optionally transformed (if not transformed, the ReadData() has a parameter to apply log2 transformation). Sample meta-data ('sampleMetaData') must include at least one phenotype column (to calculate the linear model interaction term). Meta-data for genes and metabolites (e.g. names, alternate ids, associate pathways) are optional. The gene expression data, gene meta-data, and sample meta-data are input into an ExpressionSet object. The metabolite abundance data, metabolite meta-data, and sample meta-data are input into a new MetaboliteSet object, a new eSet object designed to contain metabolomics data (2). Both objects are integrated into a MultiDataSet object- a multi-'omics object allowing integration of eSet objects from different 'omics data sets.

**Step2: Filter/Observe Data**

The FilterData() function filters data by percentile of gene expression/metabolite abundances as well as by percent of missing or imputed metabolite values.

The ShowStats() function summarizes the metabolite and gene expression data by number of genes, metabolites, and samples for each set as well as common samples. The PlotDistribution() function allows users to produce a boxplot of the distribution of gene expression and metabolite abundance data (Figure 1).

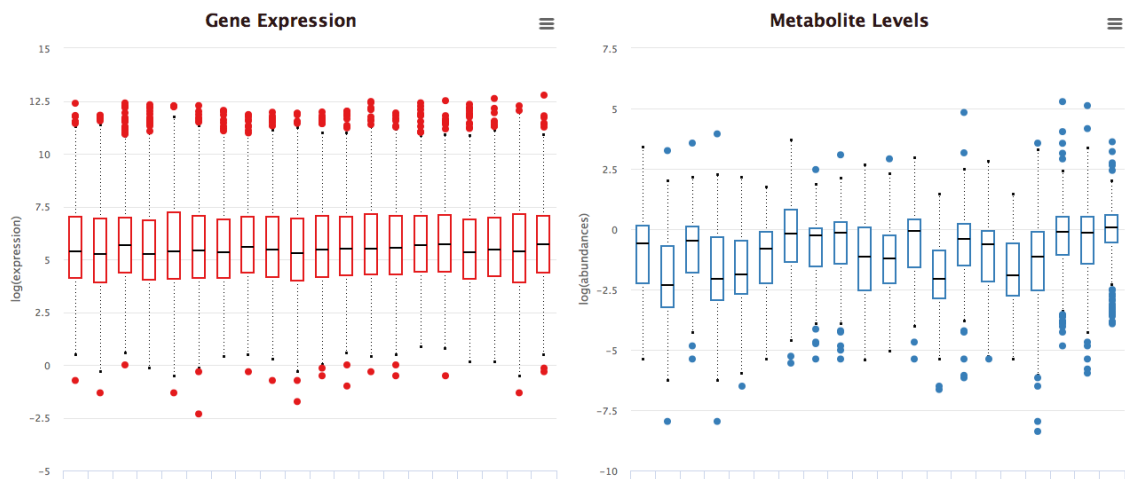

**Figure 4: Example distribution of genes and metabolites**

Prior to running the model, the user can also perform a principal component analysis of the gene expression and metabolite data using the PlotPCA() function (Figure 2). The 'stype' command allows the user to select a column from the sample meta data that color-codes the samples into two categories (two cancer types, tumor vs. non-tumor, etc).

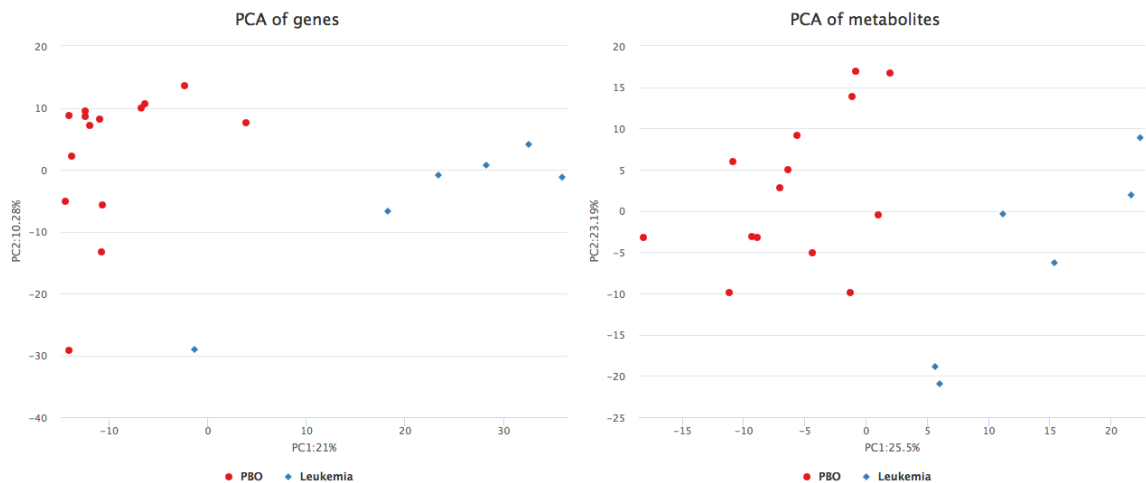

**Figure 5: Example principal component analysis plots**

### Step3: Run IntLIM

The linear models are run by the RunIntLim() function. The 'stype' command allows the user to select a column from the sample meta data for the two categories to be compared (two cancer types, tumor vs. non-tumor, etc). Currently, IntLIM only supports comparison of two categories. The resulting object from the analysis is an IntLimResults object containing slots for un-adjusted and False Discovery Rate (FDR)-adjusted p-values for the interaction coefficient. A significant FDR-adjusted p-value implies that the slope of gene-metabolite association in one phenotype is different from the other. The RunIntLim function is based heavily on the

MultiLinearModel functions developed for the ClassComparison package part of oompa (http://oompa.r-forge.r-project.org). The DistPValues() function allows the user to observe a histogram of the p-values prior to adjustment (Figure 3). The pvalCorrVolcano() function allows users to observe a volcano plot comparing the Spearman correlation difference between groups to the –log10(FDR-adjusted p-value) (Figure 4).

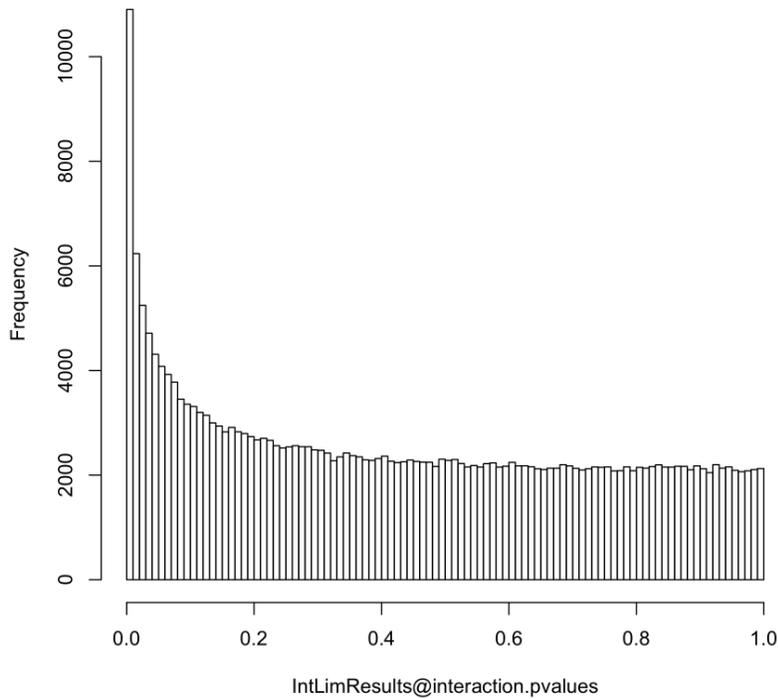

**Figure 6: Example p-values histogram**

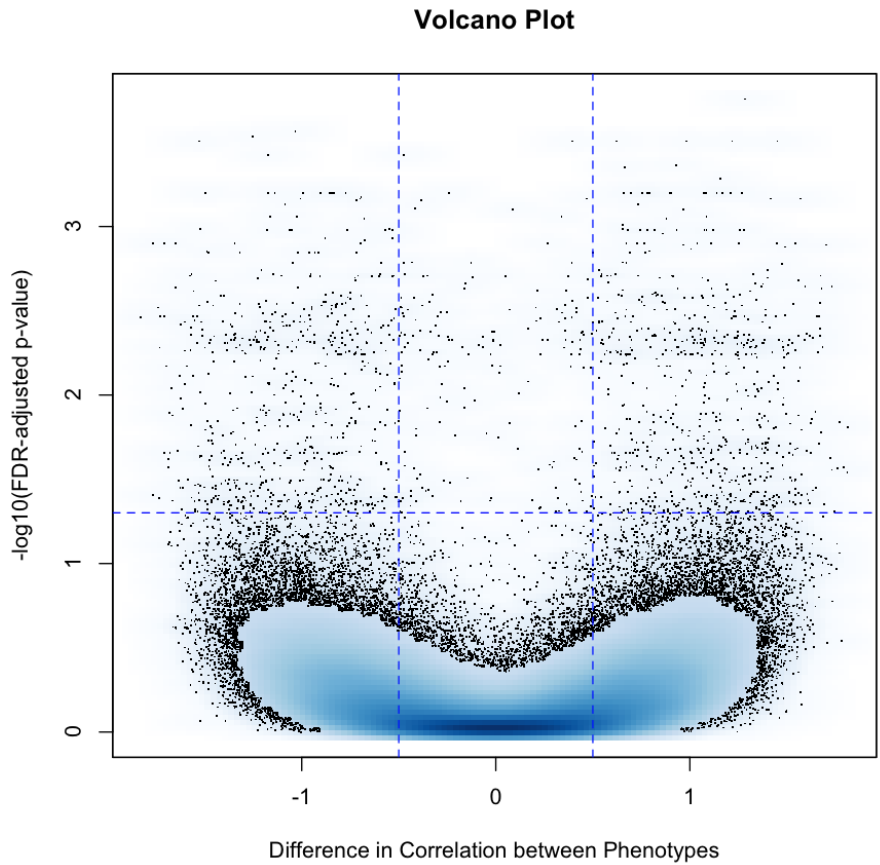

**Figure 7: Example volcano plot**

### Step4: Filter Results

The ProcessResults() function filters the results by FDR p-values (default set at 0.10) and by the absolute value difference of the gene-metabolite Spearman correlation (default set at 0.50) between the two groups. The output is a list of gene-metabolite pairs and gene-metabolite Spearman correlations for each of the two groups.

The CorrHeatmap() produces a clustered heatmap of these correlations (Figure 5).

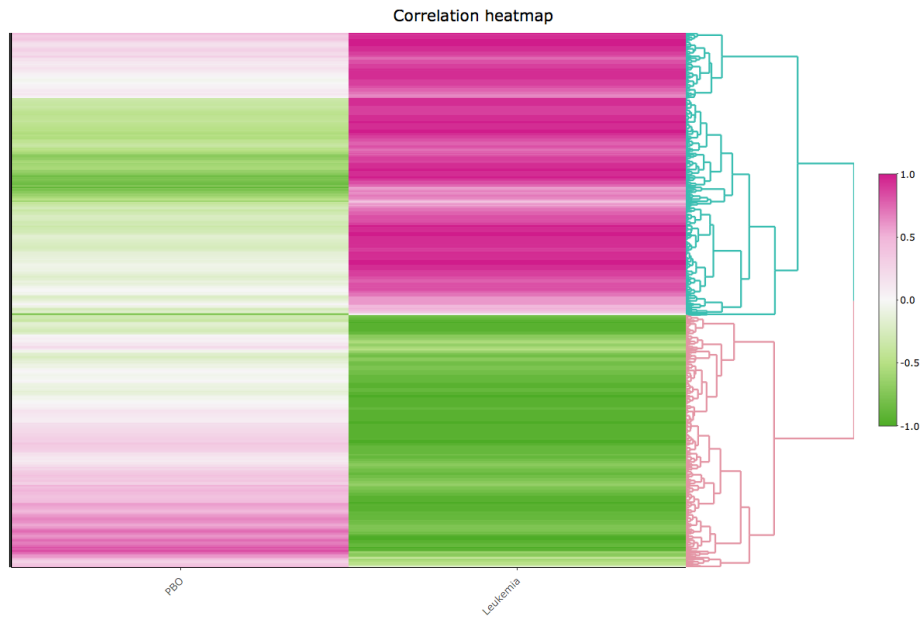

**Figure 8: Example gene-metabolite correlation heatmap**

**Step5: Visualize and Export Results**

A PlotGMPair() function produces a scatterplot of a user-selected gene-metabolite pair, which is color-coded by phenotype (Fig. 6)

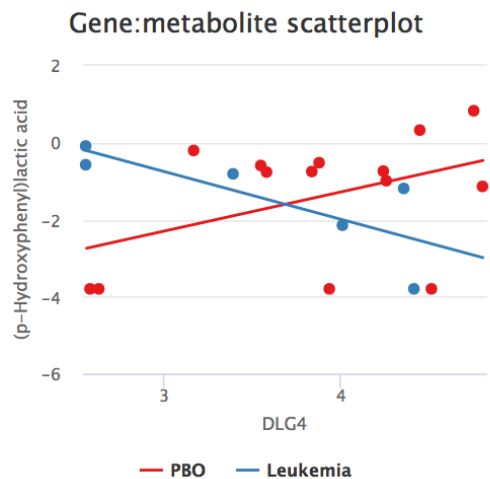

**Figure 9: Example of gene-metabolite scatterplot**

Importantly, most plots generated in IntLIM use Highcharter (http://jkunst.com/highcharter/) and Plotly (3) (https://plot.ly), which enables interactive visualization and allows users to promptly assess the effect of changing parameters on analysis results and accelerating discovery of phenotype-specific gene-metabolite pairs. This will greatly allow the workflow to be accessible to non-computational biologists.

The OutputData() and OutputResults() function produces tables of data and results of the analyses into zipped CSV files.

**ShinyApp User Interface**

A Shiny App embedded in the package provides a user-friendly interface for running IntLIM (https://shiny.rstudio.com) (Fig 7). The App calls functions from the R package so includes all the features describe above. These features include allowing users to input and observe distributions of transcriptomic and metabolomic data (Figure 7A), to filter input data (Figure 7A), to produce distribution of adjusted p-values for interaction coefficients (Figure 7B), to produce volcano plots of Spearman correlations/-log10(FDR-adjusted p-value) (Figure 7B), to produce a heatmap of gene-metabolite correlations for the two selected groups (Figure 7C), and to produce scatter-plots of select gene-metabolite pairs (Figure 7D). The App can be called by typing "runIntLIMApp()" in the R console or RStudio.

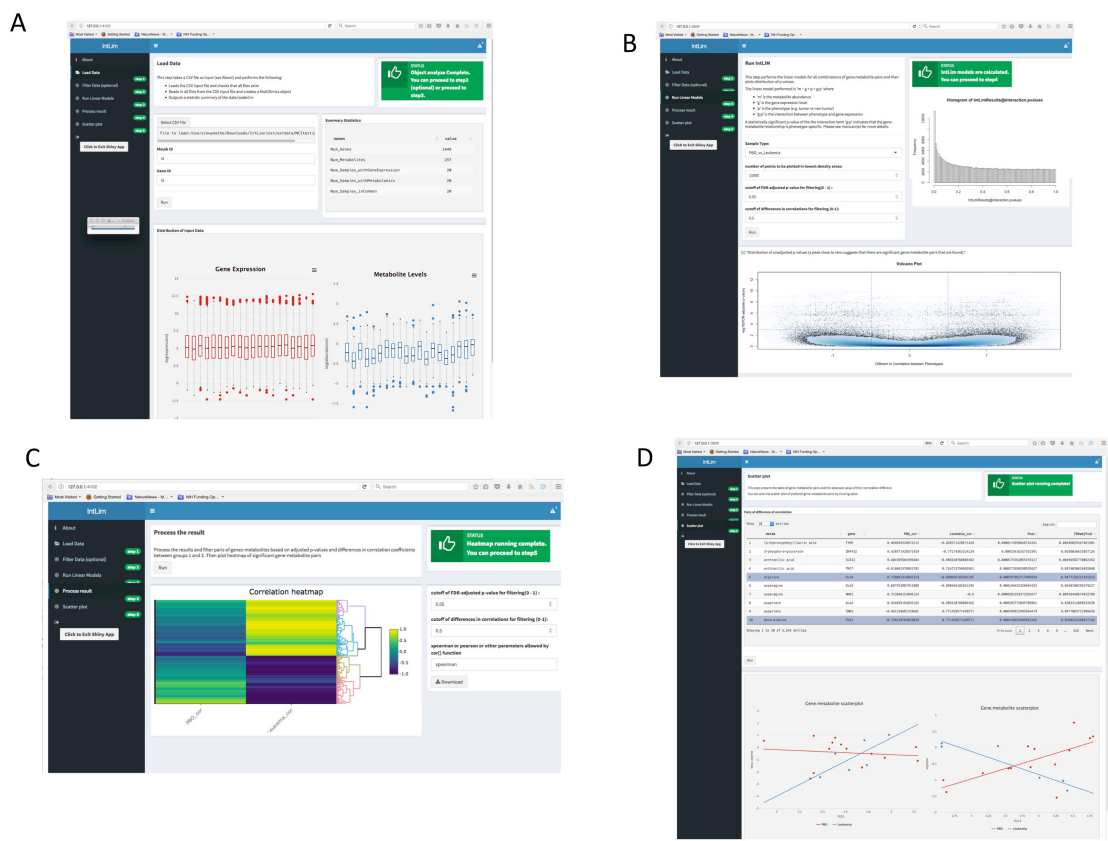

**Figure 10: Example of Shiny App. A. Inputting and filtering data. B. Running IntLIM model and observing p-value distribution and volcano plots. C. Observing heatmap of gene-metabolite correlations. D. Scatterplots of select gene-metabolite.**

## Vignettes

The IntLim Github repository (https://github.com/mathelab/IntLIM) includes a vignette with a test data set, which includes a subset of gene and metabolite levels from the original NCI-60 cell line data(4). This reduced dataset allows the user to work through the steps of the workflow. For the data analyzed in this publication, the NCI-60 cell line data with vignette is available at https://github.com/Mathelab/NCI60_GeneMetabolite_Data, and the breast cancer data with vignette is available at https://github.com/Mathelab/BreastCancerAmbs_GeneMetabolite_Data.

## Supplementary Figures

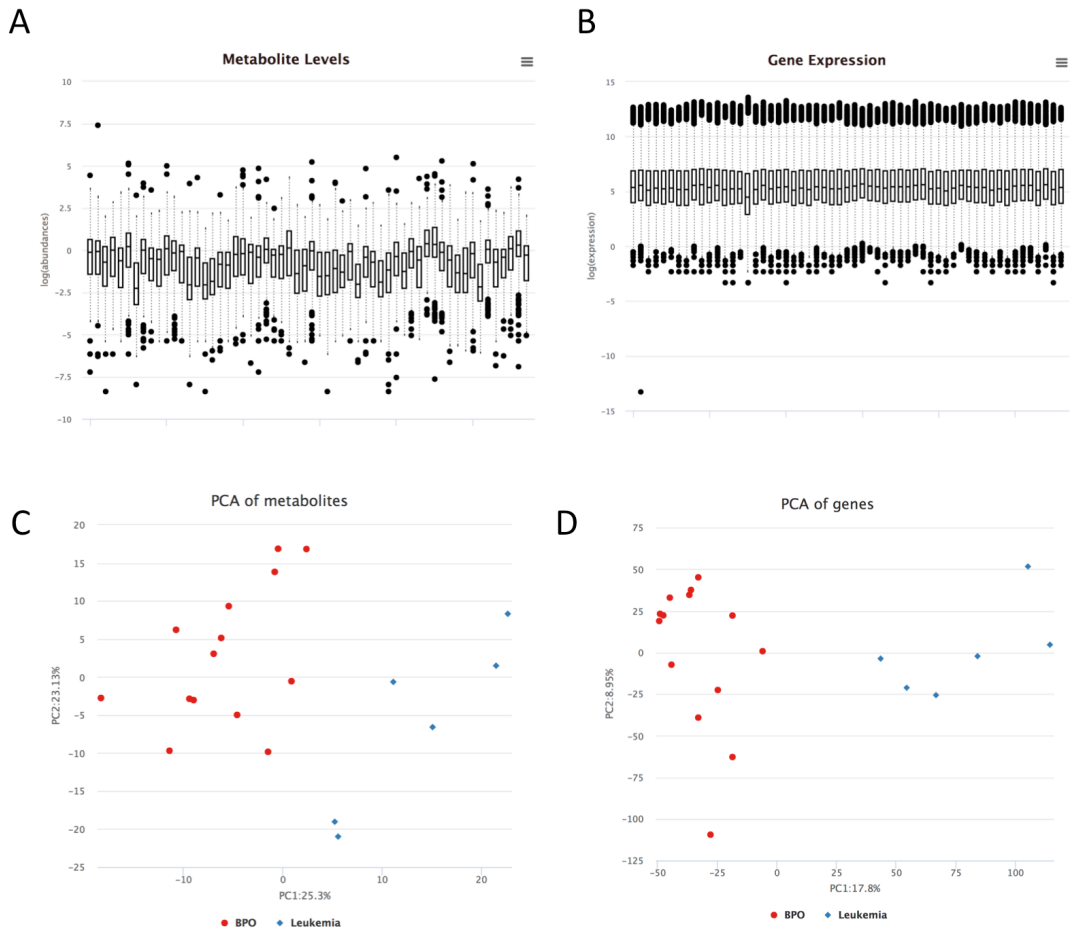

**Figure S1**: Preliminary analysis of filtered NCI-60 data involving 14 breast/prostrate/ovarian cancer (BPO) lines and 6 leukemia cell lines with 220 filtered metabolites and 16,188 genes . A) Distribution of normalized (Metabolon method) metabolite abundances among NCI-60 cell lines. B) Distribution of normalized (MAS5 algorithm) gene expression data. C, D) Principal component analysis of metabolomics and gene expression data, respectively. In the IntLIM package Rshiny app, these plots are interactive and hovering over points will provide information on those points (e.g. sample names).

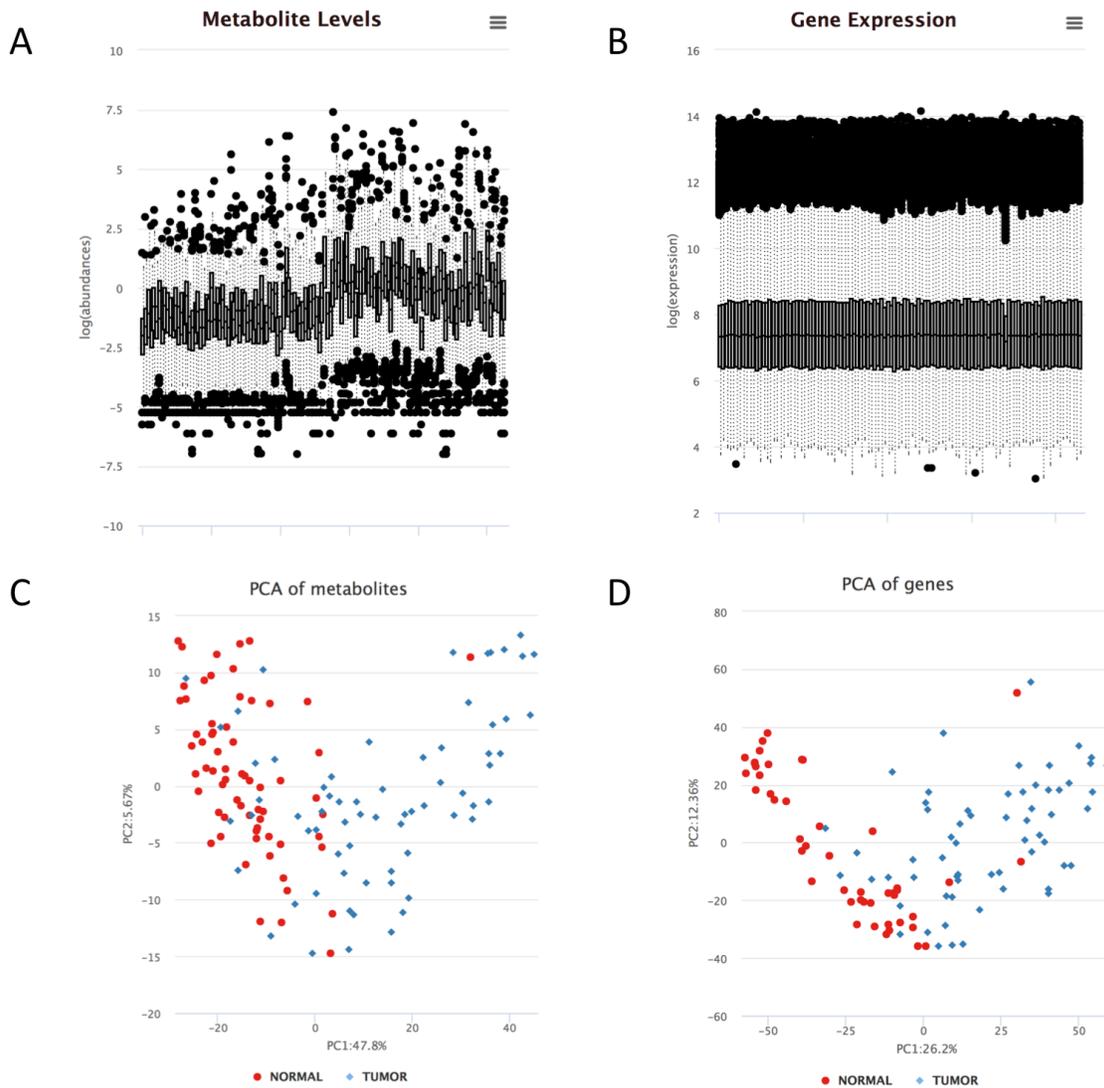

**Figure S2**: Preliminary analysis of filtered breast cancer data involving 108 samples (61 tumor and 47 non-tumor) with 379 metabolites and 18,228 genes. A, B) Distribution of normalized metabolite levels (Metabolon method) and RMA-normalized gene expression levels for all samples, respectively. C,D) Principal component analysis of metabolomics and gene expression data, respectively. In the IntLIM package Rshiny app, these plots are interactive and hovering over points will provide information on those points (e.g. sample names).

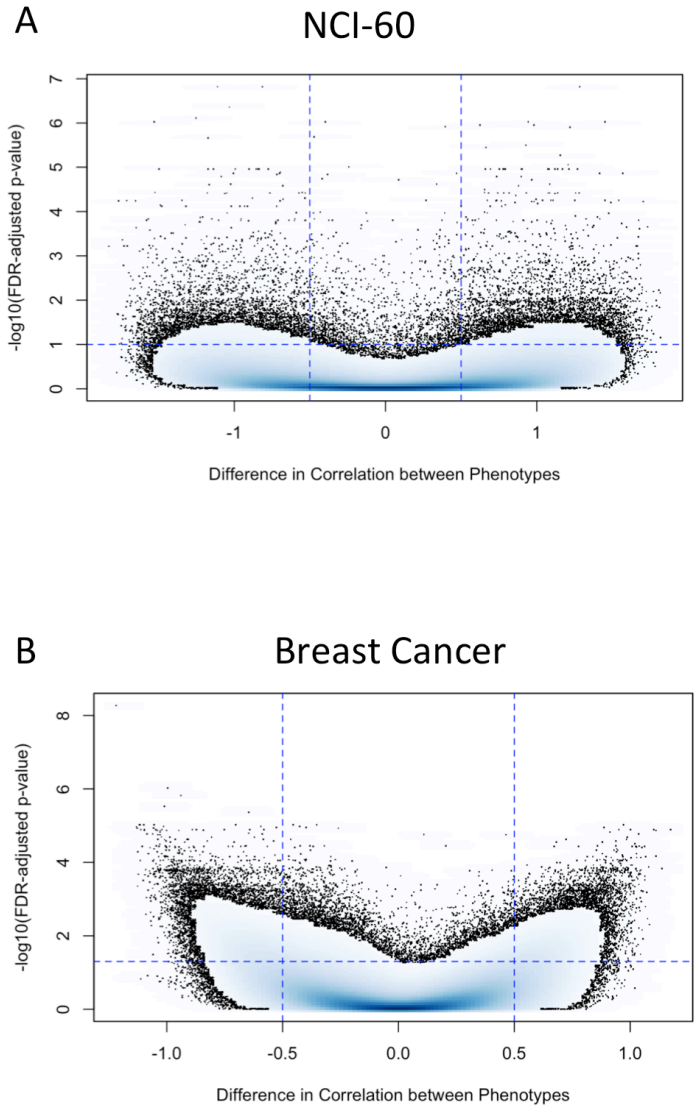

**Figure S3:** "Volcano plots" of Spearman correlation differences vs. FDR- adjusted p-values (of interaction term in linear model, see Methods) for A) NCI-60 cell line analysis and B) breast cancer data analysis.

# Supplementary Tables

**S1 Table: NCI-60 Data pathway analysis results of genes.** Ingenuity Pathway Analysis Canonical Pathways from Genes involved in Gene-Metabolite Pairs of the Leukemia Correlated Cluster and Leukemia Anti-Correlated Cluster. P-values are all calculated from right-tailed Fisher's Exact Test.

| Leukemia Correlated Cluster | | | Leukemia Anti-Correlated Cluster | | |
|---|---|---|---|---|---|
| *Pathway* | *p-value* | *Overlap* | *Pathway* | *p-value* | *Overlap* |
| Acute Phase Response Signaling | 2.21E-04 | 6.5% (11/170) | eNOS Signaling | 3.06E-04 | 5.5% (10/181) |
| 1D-myo-inositol Hexakisphosphate Biosynthesis V (from Ins(1,3,4)P3) | 9.16E-04 | 66.7% (2/3) | CREB Signalling in Neurons | 5.08E-04 | 5.2% (10/193) |
| Hepatic Fibrosis/Hepatic Stellate Cell Activation | 1.55E-04 | 5.5% (10/183) | dTMP De Novo Biosynthesis | 9.55E-04 | 21.4% (3/14) |
| CDK5 Signaling | 1.84E-04 | 7.1% (7/99) | Huntington's Disease Signaling | 1.02E-03 | 4.4% (11/249) |
| PAK Signaling | 2.06E-04 | 6.9% (7/101) | P2Y Purigenic Receptor Signaling Pathway | 1.12E-03 | 5.6% (8/143) |

**S2 Table: Breast Cancer Data pathway analysis results of genes.** Ingenuity Pathway Analysis Canonical Pathways from Genes involved in Gene-Metabolite Pairs of the Tumor Correlated Cluster and Tumor Anti-Correlated Cluster. P-values are all calculated from right-tailed Fisher's Exact Test.

| Tumor Correlated Cluster | | | Tumor Anti-Correlated Cluster | | |
|---|---|---|---|---|---|
| *Name* | *p-value* | *Overlap* | *Name* | *p-value* | *Overlap* |
| Oxidative Phosphorylation | 2.00E-15 | 16.5% (18/109) | Hepatic Fibrosis/Hepatic Stellate Cell Activation | 1.21E-06 | 8.7% (16/183) |
| Mitochondrial Dysfunction | 4.98E-14 | 11.7% (20/171) | FAK Signaling | 3.56E-05 | 10.1% (10/99) |
| Protein Ubiquitination Pathway | 5.50E-04 | 4.2% (11/265) | Actin Cytoskeleton Signaling | 7.28E-05 | 6.6% (15/227) |
| GDP-mannose Biosynthesis | 2.27E-03 | 33.3% (2/6) | Signaling by Rho Family GTPases | 1.94E-04 | 6.0% (15/248) |
| Pyridoxal 5'-phosphate Salvage Pathway | 9.01E-03 | 6.2% (4/65) | Circadian Rhythm Signaling | 5.18E-04 | 15.2% (5/33) |

**S3 Table: Breast Cancer Data pathway analysis results of metabolites.** Ingenuity Pathway Analysis Canonical Pathways from Metabolites involved in Gene-Metabolite Pairs of the Tumor Correlated Cluster. P-values are all calculated from right-tailed Fisher's Exact Test.

| **Tumor Correlated Cluster** | | | **Tumor Anti-Correlated Cluster** | | |
|---|---|---|---|---|---|
| *Name* | *p-value* | *Overlap* | *Name* | *p-value* | *Overlap* |
| tRNA charging | 1.33E-11 | 39.5% (17/43) | tRNA charging | 1.09E-11 | 41.9% (18/43) |
| Purine Ribonucleosides Degradation to Ribose-1-phosphate | 8.77E-07 | 58.3% (7/12) | Superpathway of Citrulline Metabolism | 6.83E-05 | 33.3% (8/24) |
| Purine Nucleotides Degradation II (Aerobic) | 1.71E-05 | 41.2% (7/17) | Urea Cycle | 1.20E-04 | 42.9% (6/14) |
| Guanosine Nucleotides Degradation II | 1.01E-04 | 50.0% (5/10) | Purine Nucleotides Degradation II (Aerobic) | 4.22E-04 | 35.3% (6/17) |
| Adenosine Nucleotides Degradation III | 1.78E-04 | 45.5% (5/11) | Purine Ribonucleosides Degradation to Ribose-1-phosphate | 5.50E-04 | 41.7% (5/12) |